\documentstyle[12pt,aaspp4]{article}
\lefthead{Brown, B. A. and Bregman, J. N.}
\righthead{Emission Mechanisms in X-Ray Faint Galaxies}
\begin{document}

\title{Emission Mechanisms in X-Ray Faint Galaxies } 

\author{Beth A. Brown \altaffilmark{1}}
\affil{NASA Goddard Space Flight Center, Greenbelt, MD 20771\\
Beth.Brown@gsfc.nasa.gov} 
\author{Joel N. Bregman} 
\affil{Department of Astronomy, University of Michigan, Ann Arbor, MI 48109-1090\\jbregman@umich.edu}
\altaffiltext{1}{National Research Council Fellow}

\begin{abstract}
Hot gas dominates the emission in X-ray luminous early-type galaxies, but in
relatively X-ray faint systems, integrated X-ray emission from discrete stellar-like
sources is thought to be considerable, although the amount of the contribution is 
controversial. To help resolve this issue, we examine the radial X-ray surface brightness
distribution of 17 X-ray faint galaxies observed with the ROSAT HRI and
PSPC.  We assume that the stellar contribution follows a de Vaucouleurs law
while the hot gas component follows a King $\beta$ model.  For some galaxies,
both models fit equally well, but for a number of systems, a dual component
model yields the best fit, from which, upper bounds are placed on the
stellar contribution.  Best-fit values for the stellar contribution are
inconsistent with (lower than) that suggested by Fabbiano, Gioia, \&
Trinchieri (1989) and estimated from the bulge of M31, but are consistent with 
the Forman, Jones, \& Tucker (1985) estimate of the stellar fraction in X-ray 
faint elliptical and S0 galaxies. Our results indicate an upper limit to discrete sources of
$L_X/L_B = 1.6\times10^{29} \, {\rm ergs \, s}^{-1}/L_{\odot}$.

\end{abstract}


\section{Introduction}
\label{sec:intro}

One key concern in the study of early-type galaxies, is the amount of
discrete X-ray sources relative to hot gas in X-ray faint ellipticals.  It is
generally agreed that integrated X-ray emission from stellar sources (such as
accreting X-ray binaries) is likely to be present at some fraction in all
elliptical galaxies. This contribution is low in X-ray--luminous galaxies (i.e. 
$L_X/L_B \gtrsim 3.2\times10^{30} \, {\rm ergs \, s}^{-1}/L_{\odot}$), where
hot, diffuse gas dominates the total X-ray emission (e.g. Forman, Jones \& 
Tucker 1985; Canizares, Fabbiano \& Trinchieri  1987; 
Davis \& White 1996; Brown \& Bregman 1998; 
Buote \& Fabian 1998). The fraction of X-ray emission from 
stellar sources is expected to increase with decreasing $L_X$, but the $L_X$
at which the stellar component ($L_{X,\star}$) dominates the observed emission is still debated.

Using {\em Einstein Observatory}, {\em ROSAT}, and {\em ASCA} data,
researchers have derived estimates (or upper limits) to the stellar
X-ray emission in E and S0 galaxies using various methods.
Forman, Jones \& Tucker (1985) determined an upper limit of 
$L_{X,\star}/L_B \simeq 4\times10^{28} \, {\rm ergs \, s}^{-1}/L_{\odot}$
({\em Einstein} IPC band) by assuming discrete sources make a 50\% 
contribution to the total diffuse X-ray emission of Cen A.  In this limit, only the 
faintest galaxies ($M_B > -19$ in their sample) can have their X-ray emission
dominated by discrete-source emission. Irwin \& Sarazin (1998) 
concluded that the X-ray emission of faint galaxies 
($L_X/L_B < 5\times10^{29} \, {\rm ergs \, s}^{-1}/L_{\odot}$) is a combination
of a hard stellar component and a very soft component, which is most
likely stellar in origin as well. This was inferred by comparing the X-ray ``colors"
of a sample of early-type galaxies to the bulge of M31 
($L_X/L_B = 3.2\times10^{29} \, {\rm ergs \, s}^{-1}/L_{\odot}$; Irwin 2000,
private communication). 
Fabbiano, Gioia \& Trinchieri (1989) used the average 
$L_X/L_B$ of early-type spirals 
($L_X/L_B = 4\times10^{29} \, {\rm ergs \, s}^{-1}/L_{\odot}$,
for $L_X = 10^{38} - 10^{42} \, {\rm ergs \, s}^{-1}$) as the
benchmark for the stellar contribution. By comparing the $L_X$ and $L_B$ 
of elliptical and S0 galaxies to spirals (in the {\em Einstein}
0.5--4.5 keV), they concluded that X-ray emission from hot gas is not 
significant in systems for which $L_X < 10^{41} \, {\rm ergs \, s}^{-1}$.
In the various estimates of the stellar emission fraction, there is an order of magnitude difference, making an accurate determination even more necessary.

The primary mechanisms for the X-ray emission can be determined through
the examination of X-ray spectra (e.g., 
Trinchieri, Kim, Fabbiano, \& Canizares 1994; 
Loewenstein \& Mushotzky 1997; Buote \& Fabian 
1998). A hard spectrum ($T_X \sim$ 10 keV) would indicate 
the presence of evolved stellar sources, while soft X-rays ($T_X \sim$ 1 keV) would be 
reflective of a hot gas component.  Recent work incorporating the use of
multi-temperature spectral models indicate the presence of a very soft 
component in addition to the hard and soft components described above,
the nature of which is suspected by most to be stellar in origin but which
may be due in part to a warm ISM (Pellegrini \& Fabbiano 1994; Fabbiano, Kim, \& Trinchieri 1994; Kim et al. 
1996). {\em ROSAT}, although it has better angular resolution 
than the {\em Einstein Observatory}, does not provide high enough spectral 
resolution to show signatures unique to a stellar population, especially in fainter 
systems.  Another problem with spectral modeling is that acceptable fits can 
be obtained with both single temperature models and multi-temperature models. 
Even with the better resolution of {\em ASCA}, calibration problems at low
energies may affect spectral modeling.

The use of X-ray radial surface brightness profiles offers another way to 
determine the relative fractions of gas and discrete sources to the observed X-ray
emission. Previously, these profiles existed only for X-ray-bright galaxies 
for which adequate signal-to-noise data was achieved, and were used to obtain 
gas densities and masses, test predictions of the cooling flow model, or model
the dynamical evolution of gas flows (Sarazin \& White 1988; Pellegrini \& 
Ciotti 1998; David, Forman \& Jones 1991).  The use of 
surface brightness profiles to infer the contributions to the X-ray emission has been limited, 
and few conclusions have been reached with regard to faint galaxies 
(Forman, Jones \& Tucker 1985; Trinchieri, G.,
Fabbiano, G., \& Canizares, C. R. 1986;
Pellegrini \& Fabbiano 1994). 

Data has become available through {\em ROSAT} that better allow 
for the examination of radial surface brightness profiles of fainter galaxies.  
Brown \& Bregman (1998, 2000) provide X-ray information
on a {\em ROSAT} survey of early-type galaxies extending to fainter X-ray luminosities
than available from {\em Einstein Observatory} data.  In this paper, radial surface brightness 
profiles are used to determine or set upper limits to the stellar contribution in the less
luminous galaxies of this survey.

\section{The Sample}
\label{sec:sample3}

The galaxies chosen for this study (see Table \ref{tab:sub}) are the X-ray faintest 
galaxies of the Brown \& Bregman (1998) survey of early-type galaxies.
Galaxy distances are derived from Faber et al.\ (1989) using an 
$H_0$ = 50 km/s/Mpc (the McMillan et al.\ 1994 distance is used 
for NGC 5102). Values for the stellar velocity dispersion, $\sigma$, are also obtained 
from Faber et al.\ (1989), from which the dispersion temperature, $T_{\sigma}$, is 
calculated according to $kT = \mu m_p \sigma ^2$ (where $\mu$ is the mean molecular 
weight, and $\sigma$ is the one-dimensional stellar velocity dispersion). The galaxies have 
optical luminosities ranging from log$(L_B/ L_{\odot}) =$ 10.2--11.2 (derived 
from Faber et al.\ 1989 magnitudes), except NGC 5102 whose distance 
determination indicates log$(L_B/ L_{\odot}) =$ 9.0. X-ray-to-optical luminosity 
ratios fall between $5\times10^{28}$ and $4.5\times10^{29} \rm ergs \, s^{-1}/L_{\odot}$.

{\small
\begin{table}[tbp]
\vspace{-0.3truein}
\begin{center}
\caption[Galaxy Properties]{\label{tab:sub}{}}

\begin{tabular}{lcclcr}
\multicolumn{6}{c}{Galaxy Properties} \cr
\tableline \tableline
{Name} & {Dist} & {log $\sigma$} & 
{$T_{\sigma}$} & {log$L_B$} & {log$\frac{L_X}{L_B}$} \\
{} & {Mpc} & {(km s$^{-1}$)} & 
{keV} & {($L_{\odot}$)} & 
{($\frac{erg s^{-1}}{L_{\odot}}$)} \\
\tableline
N 1344 & 28.44{$\scriptstyle \pm \phn 1.76$} & 2.204 & 0.163  & 10.66{$\scriptstyle \pm 0.06$} & $28.81^{+0.15}_{-0.21}$ \\
N 1549 & 24.26{$\scriptstyle \pm \phn 5.12$} & 2.312 & 0.267  & 10.73{$\scriptstyle \pm 0.06$} & 29.31{$\scriptstyle \pm 0.07$} \\
N 2768 & 30.64{$\scriptstyle \pm \phn 6.50$} & 2.296 & 0.248  & 10.79{$\scriptstyle \pm 0.12$} & 29.62{$\scriptstyle \pm 0.13$} \\
N 3115 & 20.42{$\scriptstyle \pm \phn 4.30$} & 2.425 & 0.450  & 10.83{$\scriptstyle \pm 0.06$} & $28.91^{+0.08}_{-0.09}$ \\
N 3377 & 17.14{$\scriptstyle \pm \phn 2.52$} & 2.116 & 0.108  & 10.21{$\scriptstyle \pm 0.12$} & $29.21^{+0.16}_{-0.19}$ \\
N 3379 & 17.14{$\scriptstyle \pm \phn 2.52$} & 2.303 & 0.257  & 10.49{$\scriptstyle \pm 0.06$} & $29.29^{+0.16}_{-0.23}$ \\
N 3557 & 47.98{$\scriptstyle \pm 10.18$} & 2.465 & 0.541  & 11.10{$\scriptstyle \pm 0.06$} & 29.51{$\scriptstyle \pm 0.08$} \\
N 3585 & 23.54{$\scriptstyle \pm \phn 4.98$} & 2.343 & 0.308  & 10.72{$\scriptstyle \pm 0.06$} & $29.12^{+0.11}_{-0.13}$ \\
N 3607 & 39.82{$\scriptstyle \pm \phn 4.84$} & 2.394 & 0.390  & 11.18{$\scriptstyle \pm 0.12$} & 29.64{$\scriptstyle \pm 0.12$} \\
N 4365 & 26.66{$\scriptstyle \pm \phn 1.42$} & 2.394 & 0.390  & 10.79{$\scriptstyle \pm 0.06$} & 29.69{$\scriptstyle \pm 0.07$} \\
N 4494 & 13.90{$\scriptstyle \pm \phn 2.94$} & 2.095 & 0.300 \tablenotemark{a} & 10.20{$\scriptstyle \pm 0.06$} & $29.08^{+0.15}_{-0.23}$ \\
N 4621 & 26.66{$\scriptstyle \pm \phn 1.42$} & 2.381 & 0.367  & 10.78{$\scriptstyle \pm 0.06$} & $29.01^{+0.14}_{-0.16}$ \\
N 4697 & 15.88{$\scriptstyle \pm \phn 3.36$} & 2.218 & 0.173  & 10.58{$\scriptstyle \pm 0.06$} & 29.55{$\scriptstyle \pm 0.06$} \\
N 5061 & 23.92{$\scriptstyle \pm \phn 5.06$} & 2.282 & 0.233  & 10.53{$\scriptstyle \pm 0.06$} & $29.01^{+0.13}_{-0.17}$ \\
N 5102 &  \phn 3.10{$\scriptstyle \pm \phn 0.30$} & 1.820 & 0.500 \tablenotemark{a} &  \phn 8.95{$\scriptstyle \pm 0.12$} & $28.75^{+0.17}_{-0.21}$ \\
N 5322 & 33.22{$\scriptstyle \pm \phn 7.04$} & 2.350 & 0.319  & 10.80{$\scriptstyle \pm 0.12$} & 29.31{$\scriptstyle \pm 0.13$} \\
N 7507 & 35.00{$\scriptstyle \pm \phn 7.42$} & 2.377 & 0.361  & 10.82{$\scriptstyle \pm 0.06$} & $29.31^{+0.20}_{-0.34}$ \\
\tableline
\end{tabular}
\end{center}
Columns 2, 4, \& 5 derived from Faber et al. 1989 values
(Distance for NGC 5102 from McMillan et al. 1994). Column 3 from Faber et al. 1989. Column 6 from Brown 1998.\\
$^a$ Adopted value for $T_{\sigma}$.

\end{table}
}

The targeted galaxies are bounded by the stellar X-ray emission estimates of 
Forman, Jones \& Tucker (1985) and Fabbiano, Gioia \& Trinchieri 
(1989). For comparison, we have converted these estimates into
{\em ROSAT} band equivalents. The Forman, Jones \& Tucker (1985) 
{\em ROSAT} 
equivalent is $L_{X,\star}/L_B = 3.6\times10^{28} \, {\rm ergs \, s}^{-1}/L_{\odot}$
obtained by assuming a Raymond-Smith thermal model with a Galactic $N_H$ column 
density of $10^{20} \, {\rm cm}^{-2}$ and $kT = 1.0$ keV.  The
Fabbiano, Gioia \& Trinchieri (1989) estimate is
the averaged best-fit linear regression for a sample of early-type
spiral galaxies. Using an average distance of 18.12 Mpc (obtained from
Fabbiano, Gioia \& Trinchieri's 1988 sample of early-type spiral galaxies), 
and spectral model parameters above, we obtain a {\em ROSAT} band equivalent of 
$L_{X,\star}/L_B = 3.6\times10^{29} \, {\rm ergs \, s}^{-1}/L_{\odot}$.
 
X-ray surface brightness profiles for the targeted sample were extracted 
out to 4--5 effective radii ($r_e$) from processed {\em ROSAT} PSPC or HRI data. 
Observations from both instruments exist only for NGC 1549 and NGC 5322. 
Details for the general data processing of these galaxies are given in Brown \& 
Bregman (1998).  A background, taken at large radius from the 
galaxy center (typically at $> 7 r_e$), was subtracted from the data.  
Normalized PSPC and HRI data were blocked into in 4$\arcsec$ and 2$\arcsec$ 
pixels, respectively, and subsequently binned according to the resolution 
of the instrument assuming azimuthal symmetry.  PSPC data were binned in 
25$\arcsec$ wide annuli (data for NGC 3557 was binned in 35--40$\arcsec$
widths beyond $\sim 2 r_e$ to improve signal-to-noise), and HRI data were 
binned in annuli of widths increasing from 5$\arcsec$ to 40$\arcsec$ in the outer 
regions where the surface brightness was lower.  

Five galaxies had total surface brightness counts too low to be useful, and
so were not included in further analysis. NGC 4494 and NGC 4621 were also
excluded because their detections were off-axis, which affected the shape of the profiles.
Of the ten remaining galaxies, eight are classified as elliptical and two as
S0 galaxies (Tully 1988 classification).

\section{Obtaining the Stellar Contribution}
\label{sec:profiles}

Hot, interstellar gas and discrete stellar X-ray sources are the two 
primary mechanisms suggested for X-ray emission in early-type galaxies.
Moderate to low signal-to-noise data prevents extensive and detailed modeling, 
however by separately characterizing the radial surface brightness distribution 
due to each component, we seek to reach an initial indication of the relative fraction 
of stellar X-ray sources present in X-ray faint galaxies

We use a modified King model ($\beta$ model) to parameterize the X-ray
emission attributable to hot gas. The $\beta$ model was developed initially to
describe hot gas behavior in clusters, but has been often extended to individual
elliptical galaxies (Gorenstein et al. 1978; Fabricant \& Gorenstein 
1983; Forman, Jones \& Tucker 1985; 
Trinchieri, Fabbiano, \& Canizares 1986). The $\beta$ model
takes on the form
\begin{displaymath}
S(r)_{\beta} = S_0[ 1 + (r / r_c)^2]^{-3\beta + \frac{1}{2}},
\end{displaymath}
where $S_0$ is the central brightness and $r_c$ is the core radius of the X-ray emission.
Values of $\beta \approx 0.5$ and $r_c \sim$ 2 kpc are typical in brighter 
ellipticals where $L_X \approx 10^{39}$--$10^{42} \, {\rm ergs \, s}^{-1}$ 
(e.g. Sarazin 1990; Goudfrooij \& de Jong 1995).
X-ray emission of bright early-type galaxies, modeled with this King function, 
demonstrate radial surface brightness profiles that decline more slowly 
than optical profiles at large radii (e.g., Forman, Jones \& Tucker 1985).

A de Vaucouleurs $r^{1/4}$ law is used to trace the integrated emission from
discrete X-ray sources. The discrete source emission is most likely dominated
by low-mass X-ray binaries (LMXBs) in the galaxy (Irwin \& Sarazin 1998; 
David, Forman, \& Jones 1991; Trinchieri \& Fabbiano 1985).
LMXBs come from an evolved stellar population, so their X-ray brightness distribution
should follow that of stars (Pellegrini \& Fabbiano 1994). A de Vaucouleurs $r^{1/4}$ law
has been very successful in modeling the optical surface brightness of ellipticals
(Fry et al. 1999, Burkert 1993, Hjorth \& Madsen 
1991), and so we adopt it here in the form
\begin{displaymath}
S(r)_{\star} = S_e \cdot {\rm exp} \{-7.67[(r/r_e)^{0.25}-1]\},
\end{displaymath}
where $r_e$ is the effective radius (the isophote radius containing half 
of the total luminosity) and $S_e$ is the surface brightness at $r_e$. Other models,
including King models, have been used to parameterize the stellar profiles of early-type
galaxies. This was primarily because there is no straightforward transformation from
the de Vaucouleurs function  to an analytic form of the density distribution. However,
since our goal is not to derive a density distribution, this is of no concern.  

A fitting algorithm was developed to determine the four parameters -
$S_0$, $r_c$, $\beta$, and $S_e$ - that best represent the data according to the
models described above. The models are convolved with an instrumental 
point spread function (a Gaussian of $25''$ FWHM for the PSPC, and $5''$ FWHM 
for the HRI), which has the form
\begin{displaymath}
p(x) = \frac{1}{2 \pi \sigma ^2} e^{-x^2/2\sigma ^2},
\end{displaymath}
where $\sigma = {\rm FWHM}/2.35$. Spherical symmetry is assumed in the 
convolved profiles, which take the form
\begin{displaymath}
G(r) = \frac{1}{\pi \sigma ^2} \int^{\infty}_{0} \int^{\pi}_{0} e^{-x^2/2\sigma ^2} g(s) \: d\phi \: x \: dx.
\end{displaymath}
The function $g(s)$ is the true profile (gaseous or stellar), where 
\begin{displaymath}
s^2 = r^2 + x^2 - 2xr{\rm cos} \phi\:.
\end{displaymath}
The convolved profiles are optimized within the program by the $\chi ^2$ 
test, which yields the best-fit values for the four parameters.  The
optimization utilizes the downhill simplex method in multidimensions of 
Nelder \& Mead (1965).  

Surface brightness data for the targets were initially fitted to the $\beta$ model
and then the de Vaucouleurs $r^{1/4}$ law, out to 4--5 $r_e$. 
Generally, $r_c$ was held as a fixed parameter since it often cannot
be resolved by the PSPC. If adequate HRI data existed, then $r_c$
was determined by the $\beta$ model, the value of which was subsequently
used in fits to the PSPC data. If useful HRI data was not available,
a core radius of $r_c = r_e / 10$ was adopted based upon examination of
HRI data for bright elliptical galaxies (see Table \ref{tab:rcore}).
 
{\small
\begin{table}[tbp]
\begin{center}
\caption[Core Radii of Bright Elliptical Galaxies] {\label{tab:rcore}}

\begin{tabular}{lccccc}
\multicolumn{6}{c}{Core Radii of Bright Elliptical Galaxies } \cr
\tableline \tableline
{} & {} & {} & \multicolumn{3}{c}{$\beta$-model}\\
\cline{4-6}
{Name} & {$r_e / 11$} & {} & {$r_c$} &
{$\nu$} & {$\chi_{\nu}^2$}\\
{} & {($\arcsec$)} & {} & {($\arcsec$)} & 
{} & {} \\ \tableline

N 1395 & 4.10 &  &4.56 & 42 & 0.76 \\
N 1404 & 2.43 &  &5.68 & 11 & 2.84 \\
N 1549 & 4.30 &  &6.08 & 17 & 0.69 \\
N 4649 & 6.67 &  &7.66 & 30 & 1.49 \\
\tableline
\end{tabular}
\end{center}
The core radius, $r_c$, fitted within a $\beta$ model 
($\beta \sim 0.5$) and
compared to HRI data using a $\chi^2$ test. Best-fits ($\chi_{\nu}^2$) 
are for $n$ degrees of freedom ($\nu$).
\end{table}
} 

\begin{figure}[tbp]
\vspace{0.2truein}
\begin{center}
\plotone{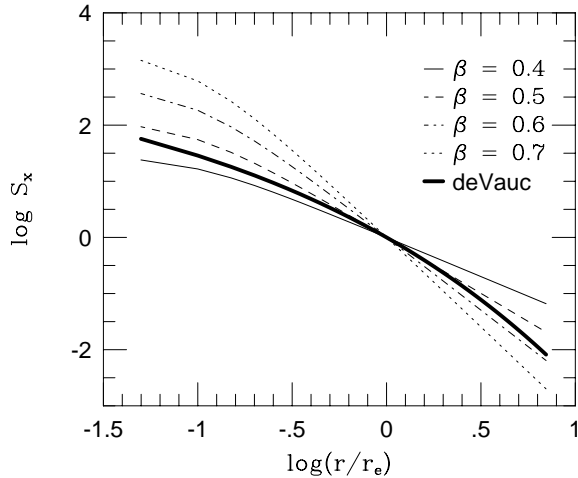}
\caption{\small A comparison of de Vaucouleurs 
$r^{1/4}$ model (heavy solid curve) and $\beta$ models for 
$\beta$ = 0.4 (light solid curve), 0.5 (dashed curve), 0.6 (dotted-dashed
curve), and 0.7 (dotted curve).
\label{fig:models}}
\end{center}
\end{figure}

Results of the $\beta$ model and de Vaucouleurs $r^{1/4}$ law fits determined 
whether a particular profile would then be fit with a two-component function where
\begin{displaymath}
S(r)_{fit} = S(r)_{\beta} + S(r)_{\star}.
\end{displaymath}
Theoretical calculations of $\beta$ models ($\beta =$ 0.4, 0.5, 0.6, and 0.7) show that 
where the diffuse X-ray emission can be described by $\beta =$ 0.5 (common in bright 
ellipticals), it is not possible to conclusively distinguish between a $\beta$ model and an 
de Vaucouleurs $r^{1/4}$ law in terms of goodness of fit (a degenerate fit, see
Figure \ref{fig:models}). A unique solution cannot be obtained with a two-component 
function to such a profile, therefore the $S(r)_{\beta} + S(r)_{\star}$ profile was
applied only when $\beta$ was determined to be steeper or flatter than $\sim$0.5

For a dual-component fit, a best-fit parameterization to the data was first obtained.
Next, upper limits to the stellar contribution (relative to the best-fit) 
were determined by incrementally increasing  $S(r)_{\star}$ until the fit became unacceptable 
at the 90\% and 98--99\% confidence levels:  
\begin{displaymath}
\chi_{limit}^2 \approx \chi _{best}^2 + \Delta \chi^2
\end{displaymath}
where $\Delta \chi^2$ = $\chi^2$ at 90\% and 99\% significance for 3 degrees of 
freedom. Once best-fit and upper limit values were found, the integrated stellar 
fraction was computed, from which log($L_{X,\star}/L_B$) can be derived. 

\section{Results of the Modeling}
\label{sec:results}

{\small
\begin{table}[p]
\vspace{-0.3truein}
\begin{center}
\caption[Single Component Galaxy Fits]{\label{tab:sin}}

\begin{tabular}{lrrrrrrrrr}
\multicolumn{10}{c}{Single Component Galaxy Fits} \cr
\tableline \tableline 
 {} &  {} & \multicolumn{5}{c}{$\beta$-model} &
\multicolumn{3}{c}{$r^{1/4}$-model}\\
\cline{3-7} \cline{8-10}
 {Name} &  {$r_e$} &  {$S_0$} &  {$r_c$ \tablenotemark{a} } &
 {$\beta$} &  {$\nu$} &  {$\chi_{\nu}^2$} & 
 {$S_e$} &  {$\nu$} &  {$\chi_{\nu}^2$}\\
 {} &  {($\arcsec$)} &  {($\frac{cnts}{pix}$)} & 
 {($\arcsec$)} &  {} &  {} &  {} & 
 {($\frac{cnts}{pix}$)} &  {} &  {} \\
\tableline
N 1549(H)$\star$ & 47.44 &   2.31 & 5.58 & 0.50 &  9 & 1.73 & 0.039 & 11 & 1.44 \\
N 1549(P)$\star$ &          &   6.38 & 5.58 & 0.46 &  7 & 3.36 & 0.150 &  8 &  3.12 \\
N 2768(P)           & 49.44 &   1.44 & 4.94 & 0.41 &  7 & 0.20 & 0.039 &  8 &  1.75 \\
N 3115(P)$\star$ & 32.32 &   5.57 & 3.23 & 0.47 &  4 & 0.90 & 0.095 &  5 &  0.80 \\
N 3379(H)           & 35.19 &   6.86 & 3.43 & 0.64 &  5 & 0.75 & 0.035 &  7 &  4.09 \\
N 3557(P)           & 37.10 & 19.59 & 3.71 & 0.56 &  4 & 0.91 & 0.160 &  5 &  4.68 \\
N 3585(P)$\star$ & 38.04 &   1.70 & 3.81 & 0.48 &  5 & 0.79 & 0.027 &  6 &  0.55 \\
N 3607(P)           & 65.49 &   4.80 & 6.55 & 0.39 & 11 & 3.03 & 0.150 & 12 & 17.10 \\
N 4365(P)           & 56.57 &   4.17 & 5.65 & 0.41 &  8 & 1.84 & 0.110 &  9 &  6.90 \\
N 4697(P)           & 73.51 &   9.20 & 7.35 & 0.40 & 12 & 3.53 & 0.260 & 13 & 22.81 \\
N 5322(H)           & 34.76 &   2.82 & 3.92 & 0.61 &  6 & 0.77 & 0.023 &  8 &  1.46 \\
N 5322(P)           &          & 19.17 & 3.92 & 0.48 &  4 & 2.94  & 0.340 &  5 &  1.65 \\
\tableline
\end{tabular}
\end{center}
Best-fits ($\chi_{\nu}^2 = \chi^2 / n$ degrees of freedom). In column 1, H=HRI, P=PSPC. A starred designation indicates 
both models fit equally well. \\
$^a$ Core radius, $r_c$, fixed for PSPC data at either HRI fitted 
values or $r_e/10$.

\end{table}
}

\begin{figure}[tbp]
\begin{center}
\plotone{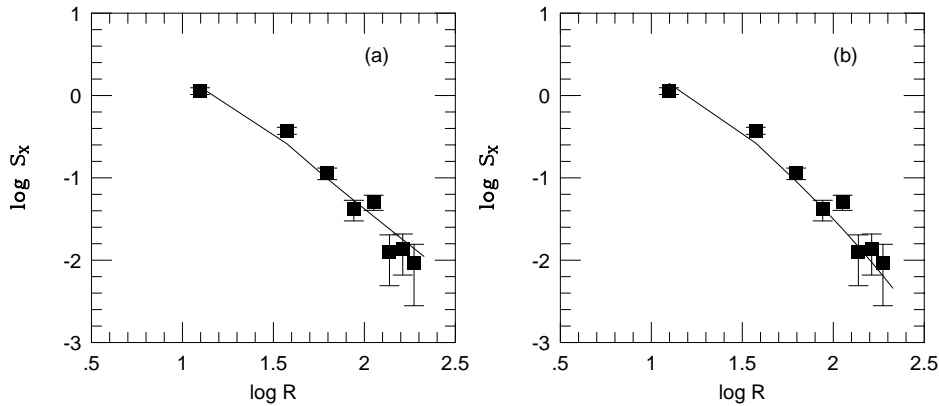}
\caption{\small NGC 1549 PSPC data fit to $\beta$ (a) and de Vaucouleurs 
$r^{1/4}$ (b) models. Surface brightness in cnts/pix is plotted against radius in arcsec. 
\label{fig:n1549}}
\end{center}
\end{figure}

Single component $\beta$ model and de Vaucouleurs $r^{1/4}$ law model fits
were performed separately to the surface brightness data for ten galaxies. A summary
of the results are given in Table \ref{tab:sin}. If a galaxy had PSPC and HRI 
data available, fits to both data sets are summarized. In addition to $r_e$, fitted 
parameter values corresponding to the minimum $\chi^2$ for both models are 
tabulated along with the number of degrees of freedom ($\nu$) and the reduced 
$\chi^2$ ($\chi _{\nu}^2$). Both $\beta$ and de Vaucouleurs models fit the 
profiles of NGC 1549, NGC 3115, and NGC 3585 equally well (see, for example 
Figure \ref{fig:n1549}), which is reflected in $\beta$ values around 0.5. Where 
$\beta$ was not 
$\sim 0.5$, pure de Vaucouleurs model fits were consistently worse compared to 
pure $\beta$ model fits, which is evidence against
complete gas-depletion in these systems. Poor fits to NGC 3607 is a result of strong 
curvature in the brightness profile, which causes the data to be underestimated in the central region and overestimated in the outer regions. Small-scale fluctuations in the outer 
brightness profile of NGC 4697 also resulted in poor fits.

\begin{figure}[tbp]
\begin{center}
\plotone{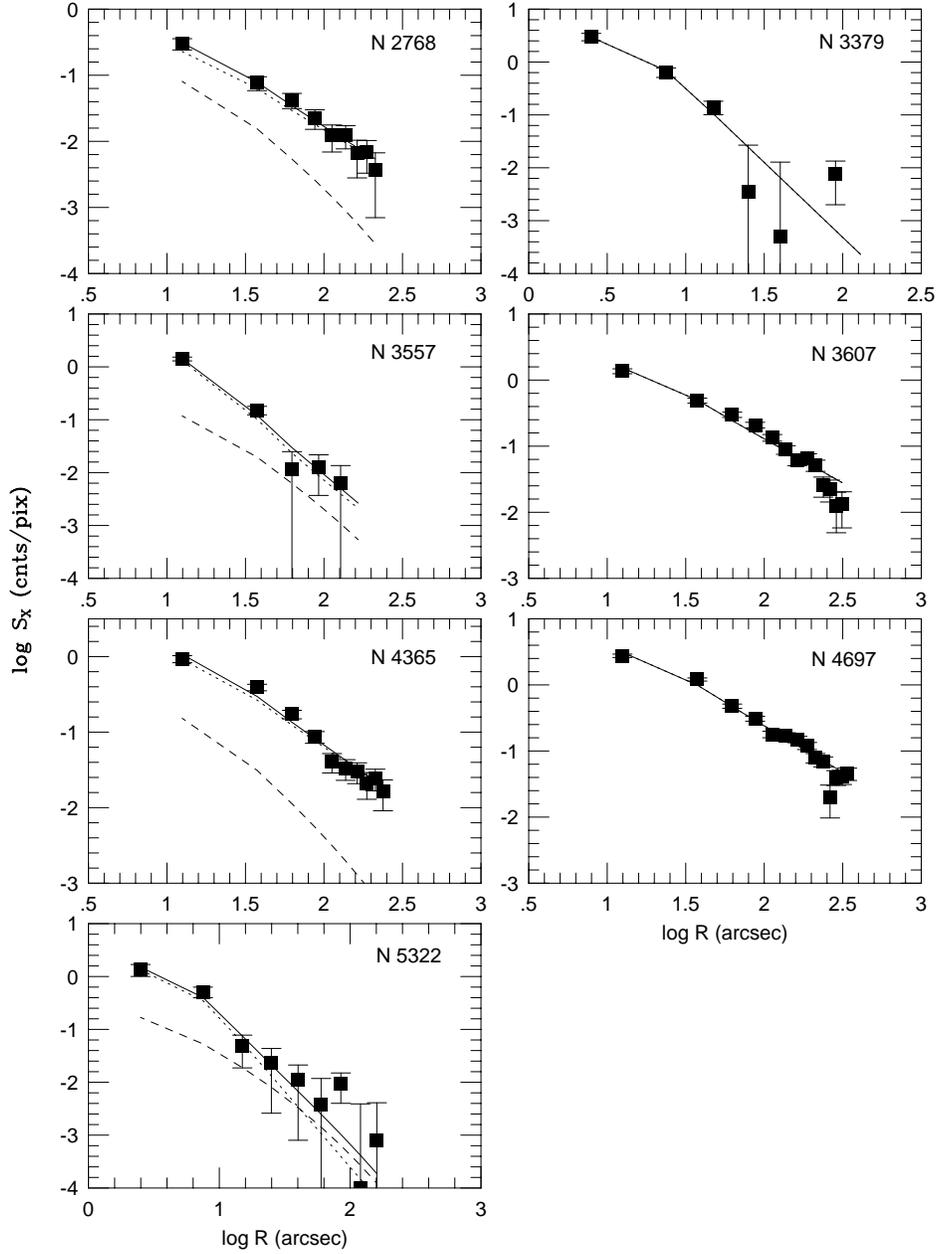}
\caption{\small Dual-component best-fits to the profiles of 
NGC 2768, NGC 3379, NGC 3557, NGC 3607, NGC 4365, NGC 4697, and NGC 5322.  In each plot, the dotted curve is the gas-component, the dashed curve is the stellar
component, and the solid curve is the combined fit. For NGC 3379, NGC 3607,
and NGC 4697 the combined fit is identical to the gas-component.
\label{fig:best}}
\end{center}
\end{figure}

The results of gas-plus-stellar component models are summarized in Table \ref{tab:dual} 
for the seven galaxies whose derived $\beta$ values were different from 0.5 in the
single-component fitting. Best-fit results as well as those for upper limits 
to the discrete source contribution (at the 90\% and 98--99\% confidence levels) are 
tabulated. The core radius, $r_c$, is fixed in all but the best-fits to HRI data.  Also 
given in Table \ref{tab:dual} is the integrated count fraction of the stellar component 
($S_{\star}/S_{tot}$), and the corresponding luminosity ratio log($L_{X,\star}/L_B$).
In the best-fit modeling, the fraction of surface brightness counts that can be attributed 
to discrete sources is less than 25\% for each of the seven galaxies modeled. The modeled profile of NGC 2768 (Figure \ref{fig:best}) indicates that the stellar component is enhanced in the central regions 
of the galaxy versus the outer regions (relative to the gas component), the reverse of what is 
indicated for NGC 3557 and NGC 5322. NGC 4365 (8\% stellar fraction) exhibits an 
approximately even ratio of stars to gas throughout the data set.  The dual-component best-fits to NGC 3379, NGC 3607, and NGC 4697 are essentially the same as in single-component
$\beta$ modeling. 

{\small
\begin{table}[tbp]
\vspace{-0.3truein}
\begin{center}
\caption[Dual Component Galaxy Fits]{\label{tab:dual}}

\begin{tabular}{lllrrlrrrr}
\multicolumn{10}{c}{Dual Component Galaxy Fits} \cr 
\tableline \tableline
{Name} & {} & {$S_0$} & {$r_c$} & {$\beta$} & 
{$S_e$} & {$\nu$} & {$\chi_{\nu}^2$ \tablenotemark{a}} & 
{$\frac{S_{\star}}{S_{tot}}$} &
{log$\frac{L_{X,\star}}{L_B}$} \\
{} & {} & {($\frac{cnts}{pix}$)} & {($\arcsec$)} & 
{} & {($\frac{cnts}{pix}$)} & {} & {} &  {} &
{($\frac{erg s^{-1}}{L_{\odot}}$)} \\
\tableline
N 2768 & best &  0.99     & 4.94 & 0.40 & 0.810E-2 & 6 & 0.23 & 0.16 & 28.83 \\
       & 90\%  & 0.24E-1 & 4.94 & 0.25 & 0.426E-1 & 6 & 1.28 & 0.79 & 29.52 \\
       & 99\%  & 0.19E-1 & 4.94 & 0.25 & 0.475E-1 & 6 & 2.12 & 0.84 & 29.54 \\

N 3379 & best &  6.87       & 3.42 & 0.64 & 0.987E-6 & 4 & 0.94 & {\tiny $<$}0.01 & 24.79 \\
       &  90\%  &  6.12    & 3.42 & 0.86 & 0.280E-1 & 5 & 1.99 & 0.68 & 29.12 \\
       &  99\%  &  5.86    & 3.42 & 0.97 & 0.352E-1 & 5 & 3.01 & 0.78 & 29.18 \\

N 3557 & best & 18.87      &  3.71 & 0.57 & 0.170E-1 & 3 & 1.20 & 0.13 & 28.63 \\
       & 90\%  &  24.39   &  3.71 & 0.99 & 0.135      & 3 & 3.25 & 0.97 & 29.50 \\
       & 99\%  &  17.45   &  3.71 & 0.99 & 0.154      & 3 & 4.94 & 0.98 & 29.50 \\

N 3607 & best &  4.79       &  6.55 & 0.39 & 0.809E-5 & 10 & 3.33 & {\tiny $<$}0.01 & 25.16 \\
       & 90\%  &  2.27  &  6.55 & 0.36 & 0.583E-1 & 10 & 3.95 & 0.24 & 29.02 \\
       & 99\%  &  1.58  &  6.55 & 0.34 & 0.783E-1 & 10 & 4.46 & 0.33 & 29.15 \\

N 4365 & best &  3.48       &  5.65 & 0.40 & 0.130E-1 & 7 & 2.09 & 0.08 & 28.61 \\
       & 90\%  &  0.43    &  5.65 & 0.31 & 0.890E-1 & 7 & 2.97 & 0.58 & 29.45 \\
       & 99\%  &  0.19    &  5.65 & 0.28 & 0.103      & 7 & 3.69 & 0.66 & 29.51 \\

N 4697 & best &   9.22      &  7.35 & 0.40 & 0.123E-4 & 11 & 3.85 & {\tiny $<$}0.01 & 25.04 \\
      &  90\%  &   4.19   &  7.35 & 0.37 & 0.103     & 11 & 4.41 & 0.27 & 28.97 \\
      &  99\%  &   2.88   &  7.35 & 0.35 & 0.136     & 11 & 4.88 & 0.35 & 29.10 \\

N 5322 & best &   2.79      &  3.88 & 0.65 & 0.434E-2 & 5 & 0.91 & 0.22 & 28.66 \\
      & 90\%  &  1.39     &  3.88 & 0.94 & 0.251E-1 & 6 & 1.65 & 0.89 & 29.26 \\
      & 99\%  & 0.50E-5 &  3.88 & 0.54 & 0.307E-1 & 6 & 2.64 & 1.00 & 29.31 \\
\tableline
\end{tabular}
\end{center}
First entry for each galaxy corresponds to ``best-fit." Second 
and third entries for each galaxy corresponds to the 90\% and 98--99\% upper 
limits to the discrete source contribution. Core radius, $r_c$, is fixed for PSPC
data, and fitted for best-fit to HRI data, and then fixed for subsequent upper limit fits. 
$\chi_{\nu}^2 = \chi^2 / n$ degrees of freedom. \\
$^a$ $\chi_{limits}^2 \approx \chi _{best}^2 + \Delta \chi^2$, where
$\Delta \chi^2$ = $\chi^2$ at 90\% and 99\% significance for 3 degrees of 
freedom. 

\end{table}
}

\begin{figure}[tbp]
\begin{center}
\plotone{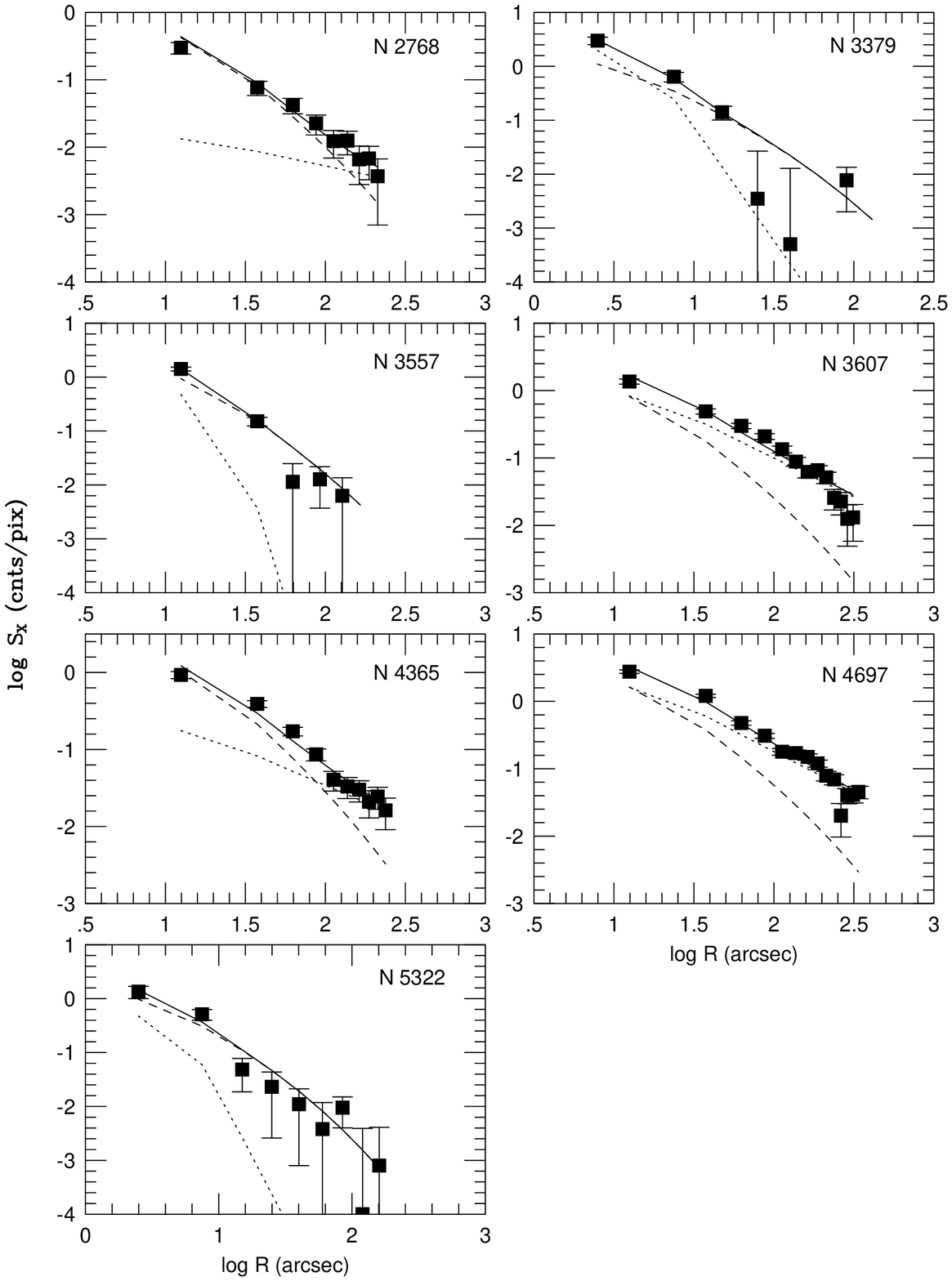}
\caption{\small Dual-component 90\% upper limit fits to the surface brightness 
profiles of NGC 2768, NGC 3379, NGC 3557, NGC 3607, NGC 4365, NGC 4697, 
and NGC 5322. Surface brightness in cnts/pix is plotted against radius in arcsec. In each plot,
the dotted curve is the gas-component, the dashed curve is the stellar
component, and the solid curve is the combined fit.
\label{fig:upper}}
\end{center}
\end{figure}

In the 90\% upper limit, the stellar fraction to the combined fit ranges from around
25\% to near 100\%. The stellar component dominates the total emission of NGC 2768
except in the outermost regions (at $> 130 \arcsec$, see Figure \ref{fig:upper}), 
and $\beta$ flattens considerably. 
In NGC 3557 and NGC 5322, a steep $\beta$ model curve rapidly falls at or before
$r_e$.  Here, the stellar component dominates throughout the radial range, however gas 
enhancement is indicated in the center. The greatest increase in stellar fraction occurs in 
NGC 3379 (from $<$1\% in the best-fit to 68\%). In this galaxy, the gas component is 
dominant only in the very center (radius $< 5  \arcsec$).  NGC 3607 and NGC 4697 
results indicate that the stellar component contributes equally, or possibly exceeds, in 
the very center, with gas dominating the total emission throughout the remainder of the 
data set. The results from NGC 4365 show stars dominating in the inner regions of the 
galaxy, with gas being more enhanced beyond radius $\sim 90 \arcsec$.

\section{Discussion}
\label{sec:disc}

\begin{figure}[tbp]
\begin{center}
\plotone{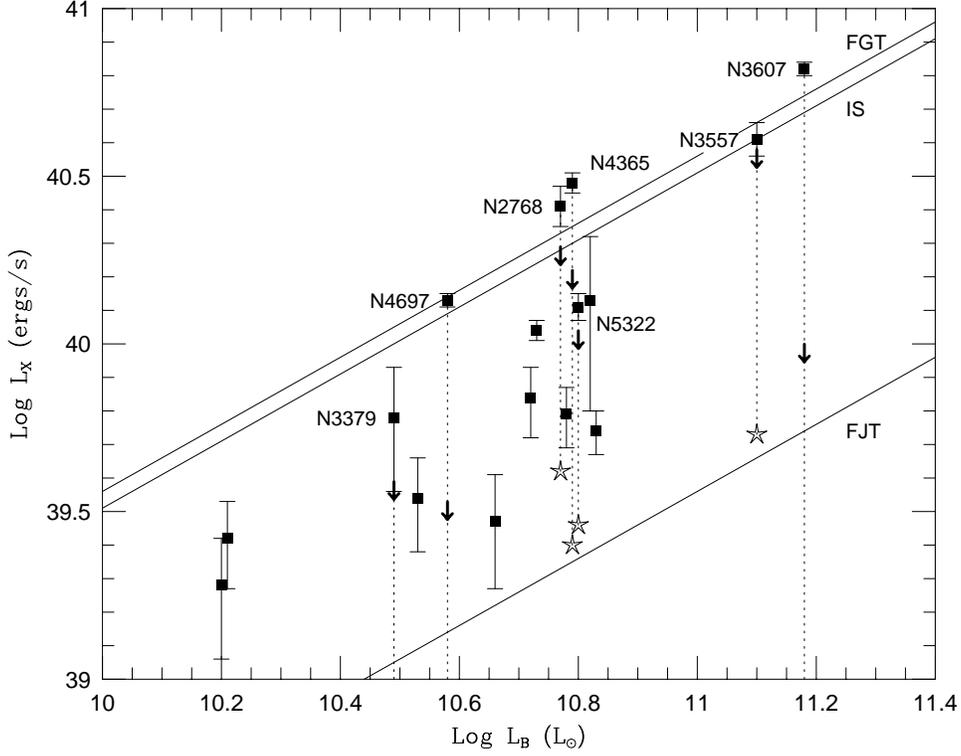}
\caption{\small X-ray vs. optical luminosities for the
target sample. Errors in $L_X$ are due to photon statistics. $L_B$ for 
NGC 2768 has been artificially lowered by 0.02 in the log for clarity, and
NGC 5102 lies beyond the plot boundaries. Stars are the ``best-fit" $L_{X,\star}$ values determined from a two-component fit while downward arrows are 90\%
upper limits. The linear lines are the stellar estimates of Forman, Jones \& Tucker 
(1985, FJT) and Fabbiano, Gioia \& Trinchieri (1989, FGT), and estimated from 
M31 (Irwin 2000, private communication) with log($L_{X,\star}/L_B$) = 28.56, 
29.56, and 29.51 respectively ($L_{X,\star}$ in ergs  s$^{-1}$, $L_B$ in $L_{\odot}$).
\label{fig:answer}}
\end{center}
\end{figure}

We compare the modeled estimates of the stellar fraction to the estimates of
Forman, Jones \& Tucker (1985, FJT) and Fabbiano, Gioia \& Trinchieri 
(1989, FGT), and that estimated from M31 (Irwin 2000, private communication).
The linear curves are superimposed upon a plot of the X-ray luminosities of 
the targeted galaxies against their optical blue luminosities (Figure \ref{fig:answer}). 
For log($L_B/ L_{\odot}) = 10.4-11$, the total $L_X$ of our sample spans the 
region between these estimates. The seven galaxies, whose radial surface 
brightness profiles could be fitted with combined King and de Vaucouleurs models, 
all have total $L_X/L_B$ values in the brighter half of the region of interest. The
best fit and 90\% upper limit $L_{X,\star}$ values are plotted for those galaxies
as well.

Best-fit modeling indicates a hot gas-dominated emission in each of the seven 
galaxies fit with the combined profile. The median for the best fits is 
log($L_{X,\star}/L_B$) = 28.61 ($L_X$ in ergs s$^{-1}$ and $L_B$ in $L_{\odot}$), 
in reasonable agreement with the
Forman, Jones \& Tucker (1985) upper limit of 
log($L_{X,\star}/L_B$) = 28.6. In the 90-99\% upper limit, however, all but 
two of the seven galaxies can be modeled with a combination profile indicative 
of a stellar-dominated emission.  The median for the 99\% confidence fits
is log($L_{X,\star}/L_B$) = 29.31, which is a factor of $\sim$1.8 and 1.6
below the limits of Fabbiano, Gioia \& Trinchieri (1989) and
M31 respectively. We further find $3\sigma$ upper limits
of log($L_{X,\star}/L_B$) = 29.15 - 29.26 for the X-ray faintest galaxies
not modeled, assuming that all of the observed emission is stellar.  We
suggest, then, an upper limit to log($L_{X,\star}/L_B$) of 29.2 for the
integrated X-ray emission from discrete sources in X-ray faint early-type
galaxies. This supports the Irwin \& Sarazin (1998) suggestion
that the brightest of X-ray faint galaxies (log $L_{X,\star}/L_B < 29.7$) may 
retain a significant amount of gas with a temperature of 0.3--0.6 keV.

We explore how our results may relate to recent spectral studies of faint E and S0 galaxies.
{\em ASCA} studies of early-type galaxies confirm a hard spectral component attributed
to the integrated emission from LMXBs (e.g., Matsumoto et al. 1997), and a 
very soft component (VSC), initially detected in {\em ROSAT} data (Kim, Fabbiano, \& 
Trinchieri 1992; Kim et al. 1996). For galaxies with 
log $L_X/L_B \geq 30.0$, spectral signatures harden as $L_X/L_B$ decreases, until the 
hard component dominates the emission. The Kim et al. (1996) study of 
an X-ray faint S0 galaxy, NGC 4382, reports that the hard spectral component contributes 
one-half to three-fourths of the total X-ray luminosity, with the remainder due to the VSC.  
This is consistent with the earlier {\em ROSAT} results which indicate that the VCS and 
hard components contribute near equally to the total X-ray emission in galaxies with 
log $L_X/L_B \leq 30.0$ ($L_X$ in ergs s$^{-1}$ and $L_B$ in $L_{\odot}$). The VSC 
may be due to a warm ISM, arise from stellar sources, or be a combination of both 
(Irwin \& Sarazin 1998; Fabbiano, Kim, \& Trinchieri 1994). Our results may be indicative of the VSC having an ISM origin, 
considering that the best-fit modeling is consistent with a substantial ISM component.

In this initial analysis of X-ray surface brightness profiles, we have assumed low-mass 
X-ray binaries (LMXBs) are the primary constituents of a discrete X-ray source population.  
LMXBs contribute to the bulk of X-ray emission in large spiral bulges such as M31 
(Trinchieri \& Fabbiano 1985). Because spiral bulges are very similar to 
elliptical galaxies in their properties and stellar populations, it is reasonable to presuppose that 
LMXBs are also the main contributors to the discrete source population in early-type galaxies 
(David, Forman, \& Jones 1991; Irwin \& Sarazin 1998). 
However, it is possible that globular clusters may be a significant source of X-ray emission 
in X-ray faint E and S0 galaxies (Trinchieri \& Fabbiano 1985), in which 
case the X-ray radial brightness distribution due to non-gaseous sources may be more 
appropriately described by a model other than a pure de Vaucouleurs $r^{1/4}$ law.

The goal of finding the stellar contribution in X-ray faint galaxies has
been hampered by low photon counts and lack of data from both {\em ROSAT}
detectors. Long observations with Chandra will be especially 
helpful in constraining the core radii in faint galaxies, fixing the shape
of the radial surface brightness profiles, and resolving bright point sources. 
It is also hoped that Chandra will provide the spectral resolution needed to 
determine whether the ``very soft component" modeled by various authors 
(see \S\ref{sec:intro}) is due to a warm ISM of to discrete sources.  Additionally, 
techniques that circumvent the problem of low counts, such as defining X-ray ``colors"
(Irwin \& Sarazin 1998), will be helpful in obtaining a definite answer to
the long-standing question of dominant emission mechanisms in X-ray faint
early-type galaxies.

\acknowledgements
We would like to thank J. Irwin and J. S. Arabadjis for valuable discussion, and
the anonymous referee for helpful comments. Also, we wish to acknowledge 
the use of the NASA Extragalactic Database (NED), operated by IPAC under 
contract with NASA. B. Brown would like to acknowledge support through a NASA 
Graduate Student Researchers Program grant NGT-51408 and a National Academy of
Science research associateship NRC-9822890. J. Bregman acknowledges 
NASA grant NAG5-3247.


\end{document}